# The Limitations of Simple Gene Set Enrichment Analysis Assuming Gene Independence

Pablo Tamayo[1], George Steinhardt[2], Arthur Liberzon[1], and Jill P. Mesirov [1,2]

1. The Eli and Edythe L. Broad Institute of Massachusetts Institute of Technology and Harvard University, Cambridge, Massachusetts 02142, USA.
2. Boston University Bioinformatics Program. Boston University, Boston, MA 02215. USA.

**Abstract**

Since its first publication in 2003, the Gene Set Enrichment Analysis (GSEA) method, based on the Kolmogorov-Smirnov statistic, has been heavily used, modified, and also questioned. Recently a simplified approach, using a one-sample *t*-test score to assess enrichment and ignoring gene-gene correlations was proposed by Irizarry et al. 2009 as a serious contender. The argument criticizes GSEA's nonparametric nature and its use of an empirical null distribution as unnecessary and hard to compute. We refute these claims by careful consideration of the assumptions of the simplified method and its results, including a comparison with GSEA's on a large benchmark set of 50 datasets. Our results provide strong empirical evidence that gene-gene correlations cannot be ignored due to the significant variance inflation they produced on the enrichment scores and should be taken into account when estimating gene set enrichment significance. In addition, we discuss the challenges that the complex correlation structure and multi-modality of gene sets pose more generally for gene set enrichment methods.

**Introduction**

The availability of global transcription profiling with microarrays in the mid 1990s made possible the analysis and interpretation of the activity of the entire transcriptome to provide insights into biological function and the mechanisms of disease. Early on, it became clear that focusing on long lists of differentially expressed genes gave limited understanding of underlying pathways and processes. Several approaches to consider testing for the over-representation of gene categories rather than genes were introduced by [1,2,3,4,5,6]. In [7] we introduced a knowledge-based approach analysis method, Gene Set Enrichment Analysis (GSEA) to address this problem. Briefly, this initial approach ranked genes by their differential expression between two phenotypes, used the Kolmogorov-Smirnov statistic to score the enrichment of an *a priori* defined sets of genes that share common biological function, chromosomal location or regulation, and evaluated the significance of the score using an empirical permutation test correcting for multiple hypothesis testing. Thus, GSEA provided a prioritized list of sets of genes for hypothesis generation and further study. In this first application we identified a set of genes involved in oxidative phosphorylation with reduced expression in diabetic patients. None of these genes were down-regulated by more than 20%, but as a group their coordinate down-regulation was significant and, together with subsequent work, lead to a better understanding of the regulation of the oxidative phosphorylation pathway [8,9] as many its components turned out to be controlled by the PCG1A transcription factor, which was itself down-regulated in diabetic patients.

As soon as this original version of GSEA appeared, objections were raised to the approach [10], some of which were immediately refuted in [11], and the rest were met by our subsequent improvement of the GSEA methodology. In Subramanian and Tamayo *et al.* [12] we introduced a version of GSEA that used a correlation-weighted Kolmogorov-Smirnov statistic, an improved enrichment normalization procedure, and an FDR-based estimate of significance that collectively made GSEA appreciably more sensitive, more general, and more robust. As a result of these improvements, and the public availability of the software and companion Molecular Signatures Database (MSigDB) [www.broadinstitute.org/gsea], GSEA became a widely used method and was applied to numerous problems across many application domains. Notably, since the original release of the software and database in 2005, the number of GSEA user registrations has grown to over 33,000, and the method used and cited in more than 3,100 scientific publications. GSEA and other gene set analysis methods have also motivated the development of general statistical methodologies for large-scale inference for "sets" of variables [13,14].

The specific knowledge-based approach pioneered by GSEA is now standard practice in the analysis of gene expression data and inspired the development of a large and growing family of conceptually similar methods. For example, Huang *et al.* [15] identified at least 68 different gene set enrichment methods in their survey. A family of popular methods estimate the over-representation of Gene Ontology (GO) annotations using a hyper-geometric statistic or Fisher's exact test (e.g., GoMiner [16], FatiGO [17], GoSurfer [18], EasyGo [19], David [20]). These methods restrict consideration to the "top" of the list, and may miss more subtle signals.  They also assume gene independence and thus produce overly optimistic results [15,21,22,23,24,25].  In addition, several improvements to gene set enrichment analysis itself have been proposed.  These include those used in [26], GSA [27], SAFE [28], Catmap [29], ErmineJ [30], and SAM-GS [31], and PROPA [32].  They employ alternative ranking metrics, enrichment statistics, and several variations on significance estimation schemes. Notably, [33] demonstrates the difficulty of finding a single, optimal statistic due to the complexity, heterogeneity and multi-modal distribution of the expression levels of genes within gene sets. Other somewhat more sophisticated methods (e.g., FunNet [34], PARADIGM [35], COFECO [36]), take a network-based approach, but restrict the analysis to processes where a deeper understanding of gene-gene interactions is already available. The primary advantages of GSEA are that it only requires gene set membership information to compute enrichment scores, considers the entire ranked list of genes, and maintains the gene-gene dependency that reflects real biology. This yields a good compromise between sensitivity, performance and applicability.

Recently, Irizarry *et al.* [37] in their "*Gene Set Enrichment Made Simple*" article proposed a "*simpler*" approach to gene set expression analysis assuming gene independence and using a one-sample *t*-test to estimate enrichment. Here we will refer to their method as SEA (Simpler Enrichment Analysis). The rationale for SEA is based on their perception that that gene independence is a reasonable simplifying assumption and thus simpler parametric approaches to gene set analysis have been ignored. Both of these assumptions disregard a large body of literature where many authors have already introduced "simple" parametric methods for gene set analysis [38,39,40,41,42]. Many researchers have demonstrated the unrealistic nature and limitations of the gene independence assumption [21,22,23,24,25,43]. In addition, they criticized GSEA for its use of an empirical null distribution, which they argue is unnecessary and hard to compute; and for using a non-parametric weighted statistic that they believe inherits the lack of sensitivity of the original Kolmogorov-Smirnov statistic. The paper concludes by proposing SEA as a serious contender and argues against the use of GSEA in any of its forms.

The aim of this paper is twofold. First, we present the SEA approach and contrast it with GSEA using the statistical framework of [43]. We then carefully consider Irizarry et al.'s [37] criticism of GSEA and refute their claims by systematically comparing SEA and GSEA on a large benchmark set of 50 datasets. We show, in agreement with earlier observations, that the gene independence assumption is not realistic because gene correlations are non-trivial and produce a substantial amount of variance inflation in the global statistic that in turn produces a large number of false positives results. Second, we discuss the challenges that the complex correlation structure and multi-modality of gene sets pose for gene set enrichment methods in general and propose that future progress in gene set analysis will result from improving the resolution of the gene sets and better ways to model the complex gene set correlation structure.

**Review of SEA and GSEA**

Here we review first the SEA approach proposed in [37] and then the GSEA approach proposed in [12] using the statistical framework of Barry et al. 2008 [43]. Several other valuable and complementary statistical frameworks for gene set analysis have been introduced in recent years [13,25,44,45,46].

First we will define the quantities relevant for the analysis: the input gene expression dataset $X$ corresponds to $N$ genes and $M$ samples and contains gene expression profiles $x_{ij}$, where $i$ indicates a specific gene (row) and $j$ a specific column (sample). The relevant phenotype of interest is defined as a vector $Y$ of $M$ binary values categorizing the sample in two groups $(Y_0, Y_1)$. The gene sets are represented as $g_k$, where $k$ runs from 1 to K, the maximum number of gene sets. We define a gene set enrichment method as a two-stage procedure incorporating i) a *local* test statistic $s_i(x_{i\cdot}, Y)$ that measures the association between a gene expression profile ($x_{i\cdot}$) and the phenotype ($Y$); ii) a *global* test statistic $S_i$ to assess the degree of the gene set's enrichment, and iii) a specific null hypothesis and error rate controlling procedure to assess statistical significance and provide a final sorted gene set enrichment result list. The global statistic can be either a parametric or non-parametric function of the local test statistics corresponding to the gene set in question ("self-contained" *statistic*) and potentially all other genes (the complement or "competitive" *statistic*);

The SEA method uses as local test statistic a two-sample *t-test* statistic $t_i$ that quantifies the magnitude of differential gene expression for each gene,

$$s_i(x_{i\bullet}, Y) = t_i = \frac{E(x_{ij} \mid j \in Y_0) - E(x_{ij} \mid j \in Y_1)}{\sqrt{\sigma^2(x_{ij} \mid j \in Y_0) + \sigma^2(x_{ij} \mid j \in Y1)}} \quad (1)$$

where the expected values and standard deviations ($\sigma$) are computed in each phenotypic class ($Y_0 / Y_1$). The SEA method can also be used with modified versions of this score [47] or with a signal to noise ratio as is used in GSEA. SEA's global test statistic is a one-sample *t*-test (z-score) that is used to estimate the "enrichment" of the entire gene set,

$$S_k^z = \sqrt{n_k} E(t_i \mid i \in g_k) = \sqrt{n_k} \left( \frac{1}{n_k} \sum_{i \in g_k} t_i \right), \quad (2)$$

The SEA method assumes that these global statistics are independent and identically distributed (gene-independence), proposes a theoretical normal-theory null hypothesis,

$$H_o^{SEA, z} : S_1^z, S_2^z, S_3^z, ..., S_K^z \text{ are i.i.d and } S_k^z \sim N(0,1), \quad (3)$$

and estimates FDR q-values following the methodology of Storey 2002 [48]. Realizing that this global statistic only detects changes in location and fails to detect other more complex types of differential gene set behavior, the SEA method adds a second $\chi^2$ global statistic,

$$S_g^{\chi^2} = \frac{\sum_{i \in g_k} (t_i - E(t_i))^2 - (n_k - 1)}{\sqrt{2(n_k - 1)}}, \quad (4)$$

and, as in the case of the z-score, an associated normal-theory null hypothesis,

$$H_o^{SEA, \chi^2} : S_1^{\chi^2}, S_2^{\chi^2}, S_3^{\chi^2}, ..., S_K^{\chi^2} \text{ are i.i.d and } S_k^{\chi^2} \sim N(0,1), \quad (5)$$

with the FDR computed in the same way as for the z-score. The SEA method is therefore based on the assumption that gene-gene correlations and gene set overlaps have negligible effects and therefore both global statistics (z-score and $\chi^2$) are independent and identically distributed under normal-theory null hypotheses $H_o^{SEA, z}$ and $H_o^{SEA, \chi^2}$. The SEA null hypotheses are instances of the gene-sampling "Class 1" null hypothesis of [43]. We will say more about the applicability of this assumption to real datasets in the next section below.

Before concluding our review of the SEA method, we point out that the second global statistic ironically renders SEA as less "simple," since the potential user now has to consider two sets of results and their corresponding FDRs. Moreover, no formal procedure is specified by SEA in order to produce a final single list of results. Instead, the authors suggest choosing high scoring gene sets according to either one of the two global statistics.

The GSEA method uses a signal to noise ratio as its local test statistic,

$$s_i(x_{i\bullet}, Y) = \frac{E(x_{ij} \mid j \in Y_0) - E(x_{ij} \mid j \in Y_1)}{\sigma'(x_{ij} \mid j \in Y_0) + \sigma'(x_{ij} \mid j \in Y1)} \quad (6)$$

Where $\sigma'$ are the intra-class standard deviations thresholded from below at 20% of the class means,

$$\sigma'(x_{ij} \mid j \in Y_*) = \begin{cases} \sigma(x_{ij} \mid j \in Y_*) & \text{if } \sigma(x_{ij} \mid j \in Y_*) \geq 0.20 \times E(x_{ij} \mid j \in Y_*) \\ 0.20 \times E(x_{ij} \mid j \in Y_*) & \text{if } \sigma(x_{ij} \mid j \in Y_*) < 0.20 \times E(x_{ij} \mid j \in Y_*) \end{cases} \quad (7)$$

GSEA uses a weighted Kolmogorov-Smirnov global statistic to assess gene set enrichment,

$$S_k^{GSEA} = \sup_{i=1,\ldots,N} \left( F_i^{g_k} - F_i^{\bar{g}_k} \right); \quad F_i^{g_k} = \frac{\sum_{h=1}^{i} |s_h|^\alpha I_h}{\sum_{h=1}^{N} |s_h|^\alpha I_h}; \quad F_i^{\bar{g}_k} = \frac{\sum_{h=1}^{i} (1 - I_h)}{(N - n_k)}; \quad I_h = \begin{cases} 1 & \text{if } h \in g_k \\ 0 & \text{if } h \in \bar{g}_k \end{cases} \quad (8)$$

An important fact, not often appreciated, is that this global statistic is weighted using a power of the local statistic $|s_h|^\alpha$ (typically with $\alpha = 1$) and it is therefore much more sensitive to differences at the top and bottom of the gene list than the standard Kolmogorov-Smirnov statistic. Because this statistic cannot be expressed as a simple function of the local statistic, it posses challenges to formal analysis based on parametric modeling.

The global statistics $S_k^{GSEA}$ depend on size of gene set and therefore are not identically distributed. GSEA addresses this issue by normalizing $S_k^{GSEA}$ values to factor out the intrinsic gene set size dependence. The relevant normalization is a change of scale using the expected value of the positive (negative) null distribution statistic induced by sample permutation,

$$\zeta_k^{GSEA} = \begin{cases} \dfrac{S_k^{GSEA}}{E(S_k^{GSEA} \mid S_k^{GSEA} \geq 0)} & \text{if } S_k^{GSEA} \geq 0 \\[1em] \dfrac{S_k^{GSEA}}{E(S_k^{GSEA} \mid S_k^{GSEA} < 0)} & \text{if } S_k^{GSEA} < 0 \end{cases}, \quad (9)$$

This rescaling is motivated by the asymptotic behavior of the Kolmogorov-Smirnov statistic (for details see supplementary information in [12]). This normalization effectively puts the gene set enrichment scores on the same scale analogous to how the factor $\sqrt{n_k}$ does it for the SEA z-score. This makes it possible to define a null distribution for the GSEA global statistics assuming a null distribution $F_0^{perm}$ induced by sample permutation,

$$H_o^{GSEA,} : \zeta_1^{GSEA}, \zeta_2^{GSEA}, \zeta_3^{GSEA}, ..., \zeta_K^{GSEA} \text{ are identically distributed and } \zeta_k^{GSEA} \sim F_0^{perm}. \quad (10)$$

The null distribution $F_0^{perm}$ corresponds to the Class 2 type subject-sampling null hypothesis of [43].

Our review of both methods clearly shows that SEA and GSEA differ in two important aspects: the choice of global statistics (z-score combined with $\chi^2$ vs. weighted Kolmogorov-Smirnov) and the specific type of null hypothesis being assumed: gene-sampling/Class 1 vs. subject-sampling/Class 2.

The authors of SEA apply the method to the handful of examples used to introduce GSEA in [12] and make the following claims in favor of the gene independence hypothesis and against GSEA:
 (i) Differential gene expression scores can be assumed to be both *independent* and *normally distributed*.
 (ii) As a consequence of (i), the *simpler* gene set enrichment method (SEA) based on a one-sample *t*-test can effectively assess gene set enrichment in generic datasets. When SEA fails to find relevant gene sets it can be "fixed" by applying a second simple statistic ($\chi^2$).
 (iii) GSEA is computationally unnecessarily complicated. The complexity of using empirical null distributions in GSEA can be avoided by using theoretical (normal) null distributions. Moreover, the gene-gene independence assumption allows the adjustment for multiple hypotheses by using independent hypotheses FDR *q*-values [49];
 (iv) SEA is faster and simpler but equivalent to GSEA and therefore GSEA should not be used. SEA should be the basis of new methodologies for gene enrichment analysis;
 (v) GSEA is based on a Kolmogorov-Smirnov statistic, which is known to lack sensitivity and thus is rarely used.

We will refute the authors' claims below by studying the gene independence assumption behind the SEA method and by performing an empirical study of 50 datasets where we focus on the consequences of assuming gene independence.

**Empirical Analysis of SEA and GSEA.**

In their SEA approach Irizarry *et al.* [37] justify the *gene-independence assumption* based on the seemingly "straight" behavior of the gene expression scores in a handful of *Q-Q* plots (Figure 3, Irizarry *et al.* [37]) and conclude that "*Barring a few outliers, which are likely associated to differentially expressed genes, the assumption appears appropriate in all datasets.*" We believe this is an over-simplification and that the gene independence assumption is not appropriate in general. Here we present the results of an empirical study to systematically evaluate SEA and GSEA and assess the effect of the gene independence assumption in a representative benchmark set of 50 expression datasets, including some from the GSEA paper [12], as well as many more from GEO (Gene Expression Omnibus), the InSilico DB database of datasets [http://insilico.ulb.ac.be] and others from the literature. The complete list of these publicly available datasets is included in Supplementary Table ST1.

This benchmark set is much more comprehensive and representative of the universe of datasets that may be used in gene set enrichment analysis. Most of the datasets derive from more recently generated data than the original GSEA examples, contain expression levels for greater numbers of genes, and display a much larger variety of phenotypic distinctions.

In order to perform our comparative analysis we ran implementations of GSEA and SEA on all 50 benchmark sets and computed the corresponding results tables, including enrichment scores, *p*- and *q*-values for each gene set. In order to make a fair comparison of both methods we computed the GSEA q-values using exactly the same procedure as in SEA (i.e., computing *q*-values using the nominal *p*-values as inputs to the *q-value* R function/package [49]). For SEA, in order to produce a single score per gene set, we generated both proposed scores (*z*-score and $\chi^2$) and chose the one with smaller p-value as was suggested in [37].

First we note that SEA uniformly produces many more significant gene sets than GSEA. For example in the Pancreas dataset [50], featured in Fig. 1B, GSEA produces 121 significant gene sets, out of a total of 1,368, at the suggested threshold (q-value < 0.25). In contrast, SEA produces 570 significant gene sets at the most stringent threshold of q-value < 0.05. This number is almost 5 times more than GSEA and accounts for 42% of the total number of gene sets. Similar remarkably large results sets are produced by SEA in other datasets. This overproduction of significant results is further exacerbated in newer more comprehensive datasets with larger numbers of genes and stronger gene-phenotype correlations. SEA produces are large number of significant sets many of which we suspect are false positives due to the assumption of gene-gene independence. It is, therefore not surprising, that among the SEA sets we find many of the significant results produced by GSEA [12].

It is difficult to assess in more direct terms the specificity/sensitivity of each method and specifically the exact percentage of false positives in SEA results because in general we do not have validated "ground truth" results for any given dataset. However, in our study we will use the methodology of Gatti *et al.* 2010 [21] where for each benchmark dataset, besides the observed global statistics corresponding to the relevant phenotype, we produced results for 1,000 randomly permuted phenotypes. Because the phenotypes have been randomized there is no significant correlation structure between the class labels and the gene profiles but the gene-gene correlations are preserved. We will use the observed and random-permutation global statistics to perform two groups of analyses: i) a study of the amount of variance inflation and the ii) inflation of p-values.

*Variance Inflation.*

One of the effects of gene-gene correlations and dependency structure in microarray datasets is the increase of variance in the global statistics. For example the SEA null hypothesis assumes the global statistics are i.i.d. and normally distributed,

$$H_o^{SEA,z} : S_1^z, S_2^z, S_3^z, ..., S_K^z \text{ are i.i.d and } S_k^z \sim N(0,1). \tag{11}$$

Barry et al 2008 [43] studied the effects of gene dependency and found that for a difference of means statistic, quite similar to the SEA z-score, and gene set sizes that are small compared with the length of the gene list, the true variance of the statistic will differ from that under the i.i.d. class 1 gene-sampling null by a "variance inflation" factor $\Gamma$ that can be approximated by,

$$\Gamma = 1 + \left(n_k \rho_k^z\right), \tag{12}$$

where $\rho_k^z$ is the average correlation between the global statistic (z-score) inside a gene set. For global statistics that are linear functions of the local statistic, $\rho_k^z$ can be approximated with $\rho_k^X$, the average correlation between gene expression profiles of genes in a dataset. From eq. 12 it is evident that the variance inflation increases with the product of the average intra-gene set correlation with the size of the gene set. Therefore, a small number of positive correlations within a gene set can result in substantial variance inflation if the gene set is large enough. For SEA the assumption of gene independence implies $\rho_k^X = 0$ and therefore a variance inflation of one. Gatti et al. 2010 [21] studied about 200 real datasets and found strong variance inflation effects, roughly in the range from 1 to 6, as a result of rather modest positive gene correlations. Their results demonstrated the importance of this effect and the unrealistic nature of the gene sampling (class 1 and SAE) null hypothesis.

In order to further investigate this issue, we analyzed the variance of the distribution of SEA z-scores under 3 null distributions: sample/phenotype-sampling (class 2), gene-sampling (class 1) and SEA (theoretical normal null). In order to produce histograms we computed z-scores generated by 1,000 permutations of the samples and 1,000 permutations of the gene identifiers. Figure 1 shows the histograms of z-scores for P53 [12] and Pancreas [50] datasets from the benchmark set.

The histograms of z-scores for 1,000 permutations of the samples (shown in grey in Fig. 1) show clearly how the gene correlations noticeably increase the width of the distribution and consequently produce significant variance inflation. For example, in the P53 and Pancreas datasets the variance of the z-score distribution for the sample permutations are 2.9 and 4.12, respectively. In contrast the variances for the gene-sampling distributions are 0.96 and 1.02, which are very close to the SEA null value of 1.

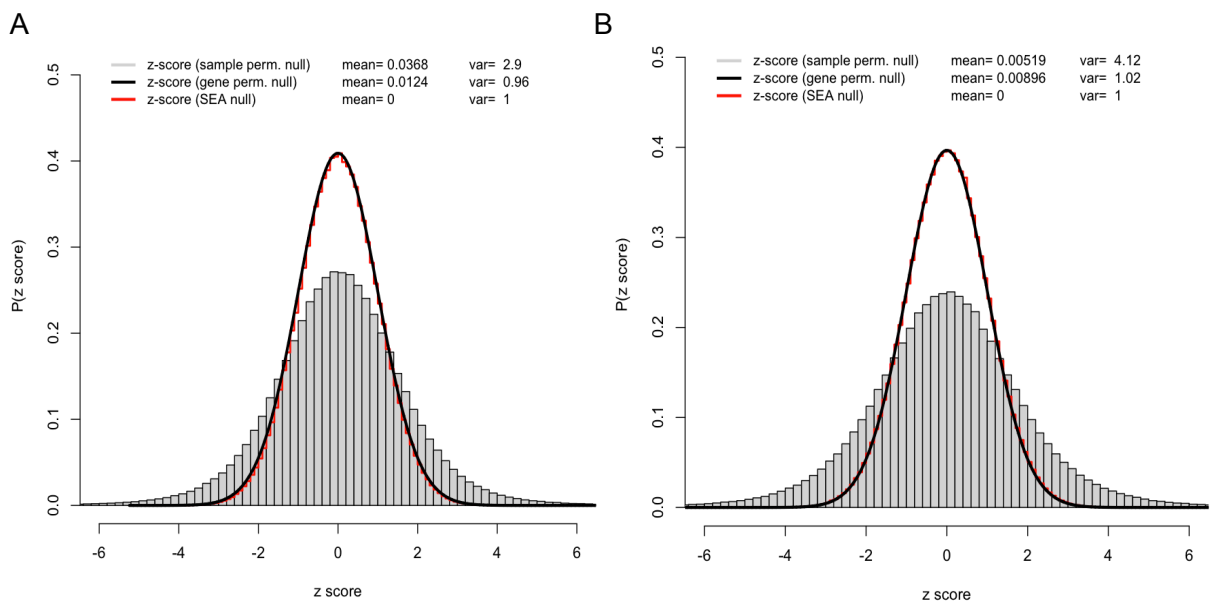

**Figure 1**. Histograms of z-scores for 1,000 permutations of the samples (grey) and gene identifiers (black), and the SEA null distribution $N(0, 1)$ for the A) P53 and B) Pancreas datasets. The legend also shows the mean and variance of the distributions.

Additional insight can be obtained from the histograms of individual average gene set internal correlations and variance inflation as shown in Figure 2. The gene correlations are on average mostly positive and relatively small, e.g., in the range [0, 0.30] (Figs 2A and 2C), but their effect on variance inflation can be significant and equivalent to a few times the variance of the SEA null distribution as estimated by eq. 12 (Figs 2B and 2D).

Because of the strong effects of the gene correlations on the global statistic distribution the sample/phenotype-sampling class 2 (GSEA) null hypothesis is a much more realistic representative of the situation found in real datasets where there is always a non-negligible amount of gene correlations.

*p-value inflation.*

The variance inflation in the global SEA statistic presented in the previous section is not only of academic interest. It has deleterious consequences in the form of high rates of false positives and inflated p-values. In Figure 3 we show histograms of *p*- and *q*-values for SEA and GSEA applied to the P53 (Fig 3A) and Pancreas datasets (Fig 3B) for 1000 random permutations of their phenotype labels. Recall that these permutations are performed to eliminate, to the extent possible, the biological differences between the two resulting groups. SEA produces a spurious over-population of low *p*-values that can be seen as a spike on the left side of the histogram in panel I of both Fig. 3A and Fig. 3B. These spikes include at least about 15% of all gene sets against the P53 dataset and about 20% of all gene sets against the Pancreas dataset.

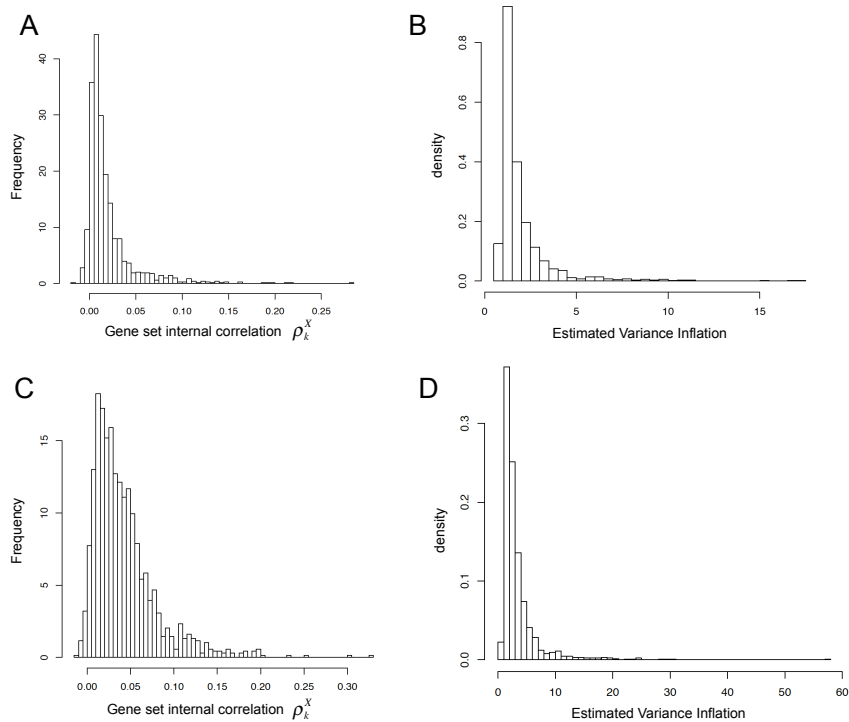

**Figure 2**. Histograms of gene correlations and estimated variance inflation for the P53 (A and B) and Pancreas datasets (C and D).

The over-population of low *p*-values translates into a significant number of gene sets with low q-values (panel II of Figs. 3A and 3B). Using a *q*-value threshold of 0.05 this spurious population includes 11.5% of all gene sets against the P53 dataset and 20.1% of all gene sets against the Pancreas dataset. In contrast GSEA produces a practically flat histogram of *p*-values and no significant numbers of gene sets with low *q*-values as expected (Figs. 3A and 3B, panels III-IV). In this case, for example, the number of gene sets with *q*-values less than 0.05 is only 0.0248% for the P53 dataset and practically zero for the Pancreas dataset. Thus we see that <u>SEA produces many significant gene sets in the absence of a biological signal as a consequence of ignoring gene-gene correlations</u>. This is a major drawback of the method.

In Figure 4 and Table ST3 we summarize the randomized phenotype results for the entire benchmark set by showing the percentage of gene sets with *q*-values less than 0.05 and 0.25 by SEA and GSEA. SEA uniformly produces a large number of false positives regardless of the choice of FDR threshold. To further demonstrate that this inflationary effect is indeed produced by the gene-gene correlations we used the same 1,000 randomly permuted phenotypes as before, but we also randomly permuted the gene identifiers effectively destroying the gene-gene correlations. In this case, SEA shows similar behavior to GSEA and neither shows inflation of p-values (Fig. 5 A-B).

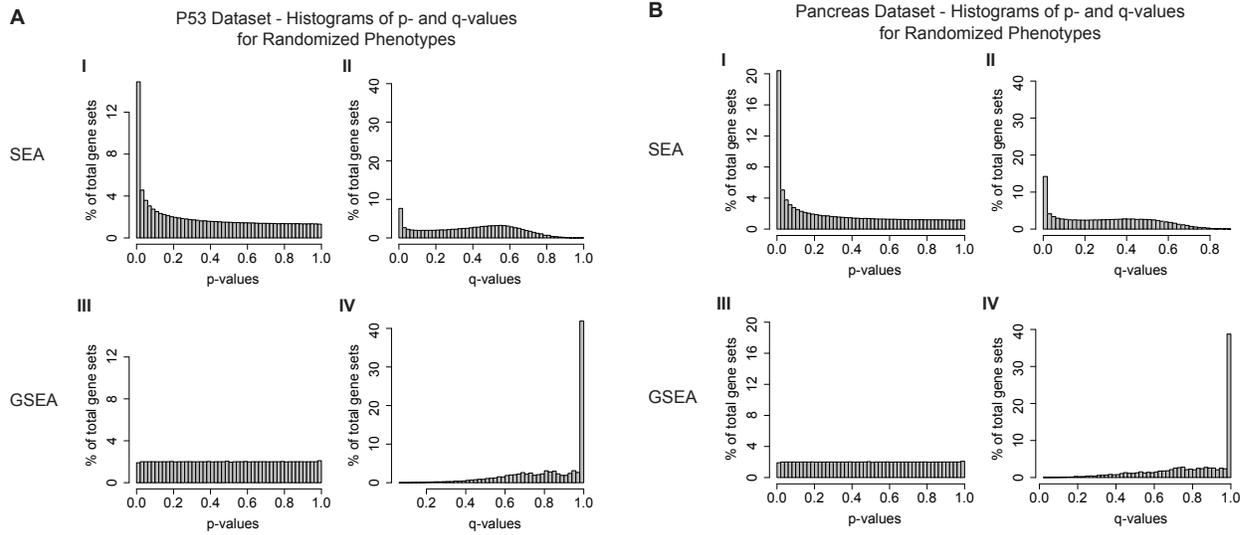

**Figure 3.** Histograms of p-values and q-values obtained by running SEA and GSEA on 1,000 randomly permuted phenotypes in the P53 dataset (A) and the Pancreas dataset (B). The y-axis shows the percentage of gene sets results.

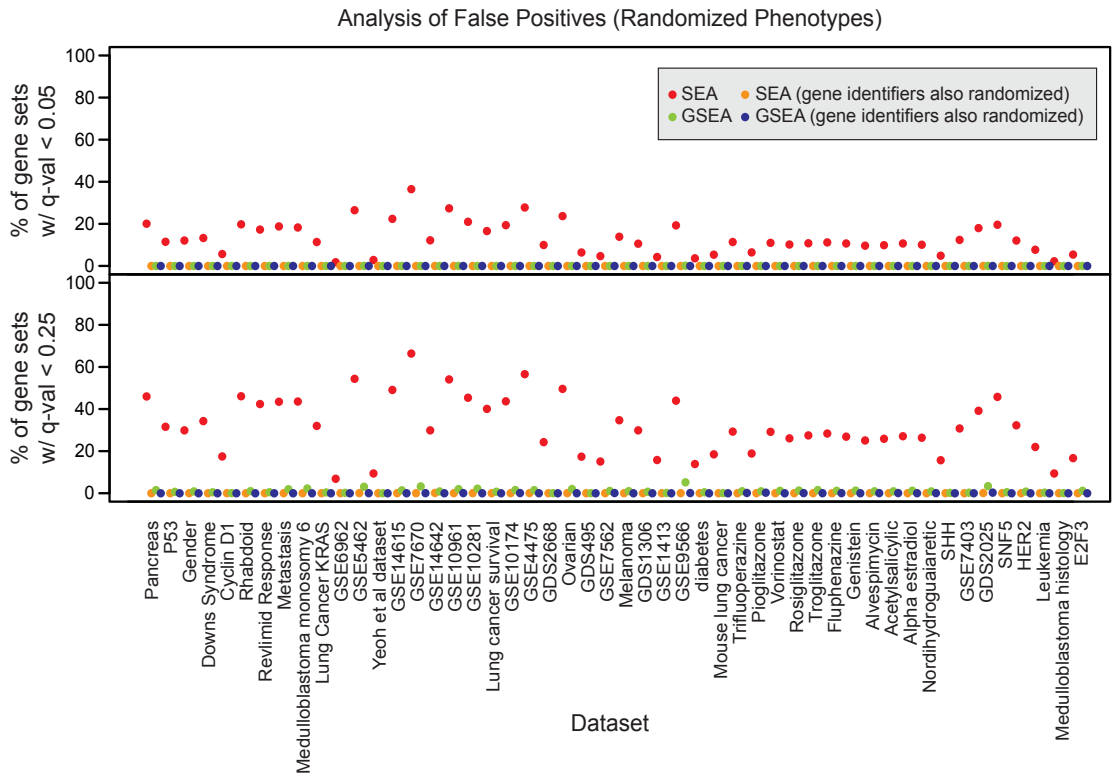

**Figure 4.** Percentage of gene sets with FDR less than 0.05 and 0.25 using SEA and GSEA in 1,000 permutations of the phenotype labels for each dataset in the benchmark set.

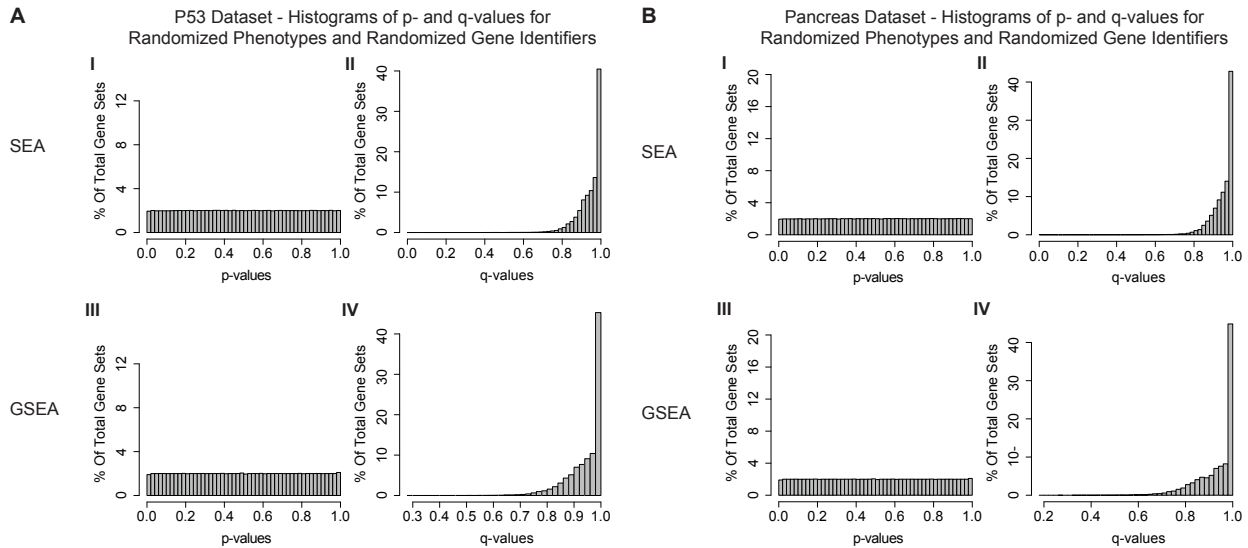

**Figure 5.** Histograms of *p*-values and *q*-values obtained by running SEA and GSEA on 1,000 randomly permuted phenotypes and randomized gene identifiers for the P53 dataset (A) and the Pancreas dataset (B). In contrast with Fig 3 here the gene identifiers have also been randomized (gene-gene correlations not preserved). The y-axis shows the percentage of gene sets results.

The above analysis convincingly demonstrates that ignoring gene-gene correlations and using the class 1(gene sampling) theoretical null distribution [13] have very negative consequences on the final results of SEA, and in fact on any enrichment method that assumes gene-gene independence. These results are very similar to those obtained by Gatti et al. 2010 [21] using different data sets and a similar, but different, statistic.

All these results cast doubt on the validity of Irizarry et al.'s claims (i)-(iv) above.

**The complexity of gene score distributions in gene sets.**

The heterogeneous distribution of the expression of genes in gene sets poses a technical challenge to properly define enrichment scores. This derives from the fact that most gene sets are noisy and imperfect, or too context-dependent to be modeled with a simple assumption or statistic. Sometimes the genes sets are derived from generic "textbook" descriptions of a biological process, e.g., Biocarta pathways or GO ontologies, and thus have little relevance to coordinately expressed components of these pathways. In other cases, they are defined in a specific cellular context different from the one in which an investigator wishes to assess enrichment or they are a mixture of multiple biological processes that may not occur coherently in any single biological sample. For all these reasons, the distribution of gene scores displays rather complex multi-modal behavior and this it turn makes it difficult to define a single enrichment score that will work well across gene sets and datasets.

Examples of this complex behavior are illustrated in Figure 6, where the behavior of three selected gene sets is shown across three of the benchmark datasets. The different panels illustrate the complex multi-modal behavior of gene sets that may occur. We note that in panels I, III, V and IX, a relatively high enrichment score of the gene set is produced by a subset of the genes, rather than the entire gene set. The genes responsible for the enrichment appear at either the top or bottom of the ranked gene list and are representatives of relevant biological processes that are indeed enriched in the studied phenotype. These examples are typical and show the complexities in modeling that gene set enrichment analyses encounter in practice. They also explain why it is difficult to describe the behavior of gene sets analytically and why overly simplistic assumptions such as those used in [37] are not likely to work. By taking these complexities into account, one can better appreciate the motivation for the weighted Kolmogorov-Smirnov statistic as a good compromise between expecting all or most genes to be coherent on one side, and overweighting one or a few genes and allowing them to dominate the score on the other. It provides a reasonable distribution-free, but empirically adaptable, way to deal with the limitations and idiosyncrasies of real life gene sets. It may not be the most powerful statistic for any given simplistic circumstance, but the modification we made in [12] of weighting by phenotype correlation is sensitive where it has to be and deals well with the behavior of real gene sets.

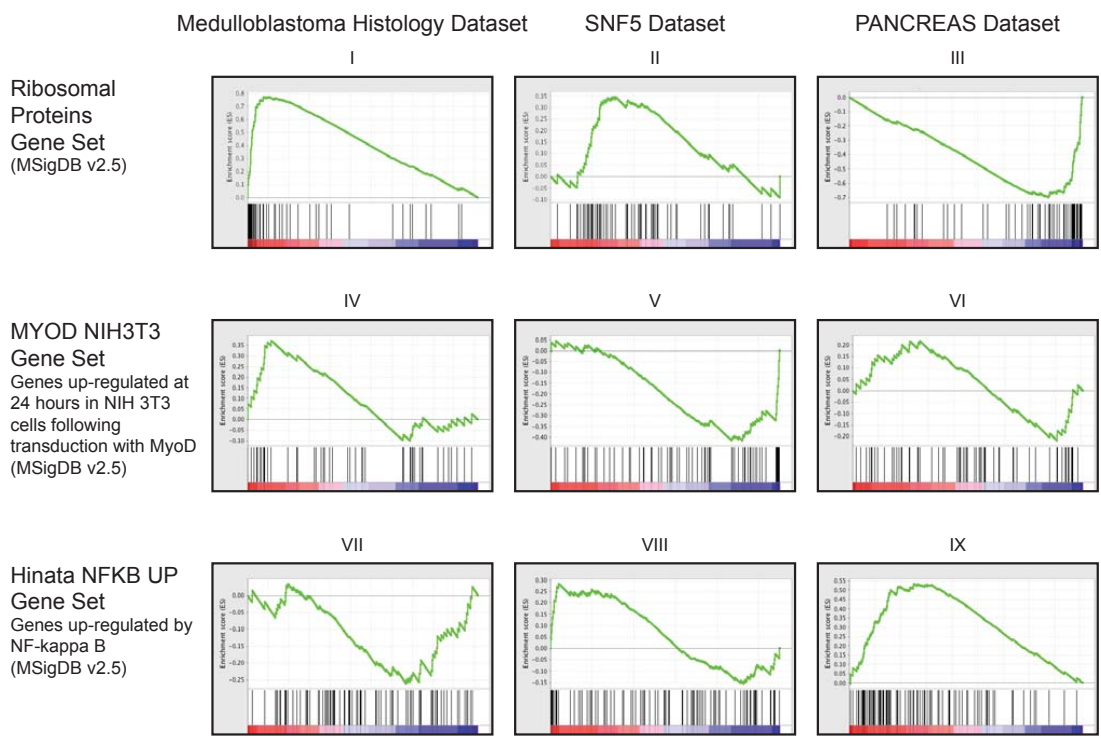

**Figure 6.** GSEA individual gene set enrichment plots: examples of top scoring gene sets that display complex behavior.

**Discussion**

The SEA method justifies the gene independence assumption based on the apparent normality of the local statistics (Q-Q plots), and the assumption that because intra-gene set gene correlations do not impact significantly the distribution of the global statistics. These assumptions are not supported by our empirical results. The strong effect of gene-gene correlations is well known and the need to take into account this dependency structure as part of gene set analysis has been well documented [22,23,24,25]. Of special note is the study of Gatti *et al.* [21] who performed a large empirical study of 200 datasets in their suggestively titled article "Heading Down the Wrong Pathway: on the Influence of Correlation within Gene Sets" and also demonstrated the practical impact of strong gene-gene correlations patterns and strongly criticized the use of the gene independence assumption.

One limitation of the SEA study [37] is that they evaluated their method, and obtained their conclusions, in only a handful of datasets, namely the examples from the GSEA paper [12]. Those datasets were good examples to illustrate the use of the recently revised GSEA method at that time, but they do not constitute a comprehensive benchmark to systematically study the properties of a new method or to perform a comparison between competing methods. The collection of gene sets they used (MSigDB in 2005) is also relatively small (522 gene sets) compared with later releases of the same collection (e.g. 1,893 gene sets in MSigDB v2.5) where the overlap between gene sets is more significant.

The authors of SEA also criticized the use of the Kolmogorov-Smirnov statistic in GSEA based on their view that it lacks sensitivity and is rarely used (their claim (v) above). However they failed to appreciate the fact that the weighted version used in GSEA is not the standard Kolmogorov-Smirnov statistic and was developed specifically to be more sensitive to differences in the tails of the distribution. Non-parametric statistics based on empirical cumulative distribution functions are the basis of new and powerful "goodness of fit" tests [51].

Irizarry *et al.* [37] also listed as one of the motivations for SEA the fact that GSEA is slow and hard to compute (claims (iii) and (iv)). In Table 1 we show typical CPU execution times (Mac book Pro 8.3) for SEA and GSEA applied to two benchmark datasets (P53 and Pancreas). As seen in the table, SEA is indeed faster than GSEA because it avoids the generation of an empirical null distribution. However, a typical GSEA run takes only about 7 minutes and for almost all users this is an acceptable running time. Moreover, SEA may be faster but this speed up comes at the high cost of large numbers of false positives as described above.

**Table 1.** Typical CPU execution times for SEA and GSEA in the P53 and Pancreas datasets (Mac Book Pro 8,3).

| Method | P53 Dataset | Pancreas Dataset |
|---|---|---|
| SEA (R implementation) | 12.32 secs. | 11.35 secs. |
| GSEA (desktop application) | 420 secs. | 444 secs. |

**Conclusions**.

We have shown strong empirical evidence that gene-gene correlations cannot be ignored and should be taken into account by gene set enrichment methods. Our results agree with the extensive literature providing theoretical or empirical evidence against the gene independence assumption [22,23,24,25] and counter the chief assumption of SEA [37].

We benchmarked SEA against GSEA in a collection of 50 expression datasets. By randomizing phenotypes, we demonstrate that gene-gene correlations produce significant variance inflation in SEA results, which also exhibit very high false positive rates and significant numbers of inflated *p*- and *q*-values. Based on our empirical results we have refuted the claims of Irizarry *et al.* [37] and more broadly we recommend that methods that ignore gene-gene correlations, such as SEA, be avoided.

Finally, we believe that the most important improvements that can be made to gene set enrichment methods are i) the improvement of gene set databases, like the MSigDB, so that their sets are less redundant and have more coherent behavior in actual biological samples; and ii) the development of more sophisticated methodologies that more accurately take into account or model gene correlations and the dependency structure in the data (i.e., the opposite of the SEA approach).

By improving the resolution of gene sets we may overcome many of the limitations we have described above: noisiness, redundancy, multiple-process representation, poor specificity, etc. We are currently experimenting with the creation of a "hallmark" collection of gene sets in MSigDB that will contain more targeted representations of biological processes. In addition, we are investigating the generation of coordinately expressed gene sets derived from the activation or repression of pathways in the laboratory.

Our group, and many others, are currently engaged in efforts to improve gene set analysis through better modeling of the datasets' correlation structure and by introducing additional information about the behavior of genes, (e.g., on a sample per sample basis and in supplementary datasets) as part of the gene set analysis. There are indeed many alternative approaches to pursue these goals and here we conclude by listing just a few: single-sample gene set enrichment [52,53,54], computational and theoretical methods for assessing size and effect of correlation in large-scale testing [22], eigenvalue-decomposition of covariance matrixes [55], rotation-based sampling [56], correlation-adjusted t-scores [57], modeling dependency among the genes within and across each gene set [58] and multiple testing procedures using dependency kernels [59]. Given the complexities of genomic data it is worthwhile to pursue new methodologies such as these; however, we should resist the temptation to over-simplify and remember the admonishments of Pearson and Tukey:

> "…it is not enough to know that a sample could have come from a normal population; we must be clear that it is at the same time improbable that it has come from a population differing so much from the normal as to invalidate the use of the 'normal theory' tests in further handling of the material."                    -- E. S. Pearson

> "Far better an approximate answer to the right question, than the exact answer to the wrong question, which can always be made precise."                    -- J. Tukey

**Methods**

Our implementation of GSEA is as described in Subramanian and Tamayo *et al.* [12]. The implementation of SEA followed the description given in Irizarry *et al.* [37]. We validated our implementation by exactly reproducing the *t*-test score and $\chi^2$ statistics from the result set of Irizarry *et al.* [37] in one of the datasets.

We obtained the differential gene-expression scores using the *signal to noise* ratio as computed in GSEA and in SEA. We also normalized the differential gene expression scores by subtracting the median and dividing by the median absolute deviation as is done in SEA

To perform the analyses of variance inflation and over-production of false positives and inflated *p*-values we followed the approach of Barry et al. 2008 [43] and Gatti *et al.* 2010 [21]. For each dataset in the benchmark we randomized the phenotype labels 1,000 times and ran both algorithms. We also computed enrichment scores by randomizing the gene labels. The *p*-values are computed using the areas under the empirical null histograms fro GSEA and areas under the normal distribution for SEA. The *q*-values were computed using the R function *qvalue* from package *qvalue* which implements the method described in Storey and Tibshirani [49].

**Acknowledgements.** We thank Helga Thorvaldsdóttir, Aravind Subramanian and other members of the Cancer Program at the Broad Institute for suggestions, and Jon Bistline for help with references and formatting the figures. This project was supported in part by grant from the National Cancer Institute (R01-CA121941).

# Supplementary Tables

## Table ST1. List of Datasets for the Benchmark Set

| # | Dataset | Number of Samples | Description | Source |
|---|---------|-------------------|-------------|--------|
| 1 | Pancreas | 36 | Pancreatic cancer vs. normal pancreas. | Abdollahi *et al.* 2007 [50] |
| 2 | P53 | 50 | NCI-60 cell lines with P53 mutant and wild type. | Subramanian and Tamayo *et al.* 2005 [12] |
| 3 | Gender | 32 | Male and Female samples. | Subramanian and Tamayo *et al.* 2005 [12] |
| 4 | Downs Syndrome | 39 | Downs syndrome samples vs. normal. | GEO dataset GSE4119 |
| 5 | Cyclin D1 | 18 | Cyclin D1 over-expression. | Lamb *et al.* 2003 [60] |
| 6 | Rhabdoid | 48 | Rahbdoid and other CNS tumors. | Jagani *et al.* 2010 [61] |
| 7 | Revlimid Response | 16 | Revlimid responders and non-responders in MDS. | Ebert *et al.* 2008 [62] |
| 8 | Metastasis | 131 | Primaries vs. metastasis. | Wolfer *et al.* 2010 [63] |
| 9 | Medulloblastoma monosomy 6 | 45 | Medulloblastoma samples with monosomy 6. | Pomeroy *et al* 2002 [64] |
| 10 | Lung Cancer KRAS | 53 | Lung cancer cell lines with mutant and wt KRAS. | Sos *et al.* 2010 |
| 11 | GSE6962 | 12 | A549 lung cancer cells treated with Ionophores vs. contr. | GEO dataset GSE6962 |
| 12 | GSE5462 | 116 | Letrozole treated vs. untreated breast samples. | GEO dataset GSE5462 |
| 13 | Yeoh et al dataset | 20 | ALL subtype E2A vs. ALL subtype TEL. | Yeoh *et al.* 2002 [65] |
| 14 | GSE14615 | 44 | Continuous remission vs. relapse in T-ALL patients. | GEO dataset GSE14615 |
| 15 | GSE7670 | 59 | Tumor vs. adjacent normal in lung cancer samples. | GEO dataset GSE7670 |
| 16 | GSE14642 | 40 | Before vs. after exercise in females blood samples. | GEO dataset GSE14642 |
| 17 | GSE10961 | 18 | Metachronous vs. synchronous liver metastasis. | GEO dataset GSE10961 |
| 18 | GSE10281 | 36 | Pre- vs. post-conventional treatment in breast cancer. | GEO dataset GSE10281 |
| 19 | Lung cancer survival | 62 | Survivors vs. non-survivors in lung cancer. | Subramanian and Tamayo *et al.* 2005 [12] |
| 20 | GSE10174 | 26 | Pre- vs. post-rasberry treatment in premalignant oral lesions. | GEO dataset GSE10174 |
| 21 | GSE4475 | 24 | Burkitt lymphoma and diffuse large B-cell lymphoma. | GEO dataset GSE4475 |
| 22 | GDS2668 | 12 | J1 embryonic cells late differentiation vs. control. | GEO dataset GDS2668 |
| 23 | Ovarian | 81 | Platinum resistance vs. sensitive ovarian cancer samples. | Dressman *et al.* 2007 [66] |
| 24 | GDS495 | 27 | VEGF-A treated vs. cntrl. umbilical cord(HUVEC) samples. | GEO dataset GDS495 |
| 25 | GSE7562 | 12 | PTEN RNA*i* knockouts vs. controls. | GEO dataset GSE7562 |
| 26 | Melanoma | 62 | BRAF mutant vs. wt short-term melanoma samples. | Lin *et al.* 2008 [67] |
| 27 | GDS1306 | 18 | MKK7D activated vs. controls in mouse heart samples. | GEO dataset GDS1306 |
| 28 | GSE1413 | 19 | mTOR inihibtion vs. controls in prostate mouse cells. | GEO dataset GSE1413 |
| 29 | GSE9566 | 18 | Astrocytes vs. neurons in mouse CNS samples. | GEO dataset GSE9566 |
| 30 | Diabetes | 34 | Diabetics vs. normal samples. | Mootha *et al.* 2003 [7] |
| 31 | Mouse lung cancer | 58 | Mouse model of Kras2-mediated lung cancer vs. normal. | Sweet-Cordero *et al.* 2005 [68] |
| 32 | Trifluoperazine | 22 | Trifluoperazine treated vs. control samples. | Connectivity Map 2.0 Lamb *et al.* 2006 [69] |
| 33 | Pioglitazone | 22 | Pioglitazone treated vs. control samples. | Connectivity Map 2.0 Lamb *et al.* 2006 [69] |
| 34 | Vorinostat | 24 | Vorinostat treated vs. control samples. | Connectivity Map 2.0 Lamb *et al.* 2006 [69] |
| 35 | Rosiglitazone | 24 | Rosiglitazone treated vs. control samples. | Connectivity Map 2.0 Lamb *et al.* 2006 [69] |
| 36 | Troglitazone | 24 | Troglitazone treated vs. control samples. | Connectivity Map 2.0 Lamb *et al.* 2006 [69] |
| 37 | Fluphenazine | 24 | Fluphenazine treated vs. control samples. | Connectivity Map 2.0 Lamb *et al.* 2006 [69] |
| 38 | Genistein | 24 | Genistein treated vs. control samples. | Connectivity Map 2.0 Lamb *et al.* 2006 [69] |
| 39 | Alvespimycin | 24 | Alvespimycin treated vs. control samples. | Connectivity Map 2.0 Lamb *et al.* 2006 [69] |
| 40 | Acetylsalicylic | 24 | Acetylsalicylic acid treated vs. control samples. | Connectivity Map 2.0 Lamb *et al.* 2006 [69] |
| 41 | Alpha estradiol | 24 | Alpha estradiol treated vs. control samples. | Connectivity Map 2.0 Lamb *et al.* 2006 [69] |
| 42 | Nordihydroguaiaretic | 24 | Nordihydroguaiaretic acid treated vs. control samples. | Connectivity Map 2.0 Lamb *et al.* 2006 [69] |
| 43 | SHH | 16 | SHH induced vs. controls in mouse cerebellum samples. | Zhao *et al.* 2002 [70] |
| 44 | GSE7403 | 19 | PDGF stimulated vs. controls in neuroblastoma cell lines. | GEO dataset GSE7403 |
| 45 | GDS2025 | 24 | Early vs. late MYC activation in pancreatic islet beta cells. | GEO dataset GDS2025 |
| 46 | SNF5 | 18 | SNF5 deficient vs. controls mouse samples. | Isakoff *et al.* 2005 [71] |
| 47 | HER2 | 21 | HER2+ vs. HER2- breast cancer samples. | Wang *et al.* 2005 [72] |
| 48 | Leukemia | 20 | B- vs. T-Cell Leukemia samples. | Ross *et al.* 2004 [73] |
| 49 | Medulloblastoma histo. | 34 | Classic vs. desmoplastic medulloblastoma samples. | Pomeroy *et al.* 2002 [64] |
| 50 | E2F3 | 19 | E2F3 induction vs. controls in breast epithelial cells. | Bild *et al.* 2006 [74] |

Table ST2. Percentage of gene sets with q-values less than 0.05 and 0.25 produced by SEA and GSEA using 1000 randomized phenotypes.

| # | Dataset | SEA Percentage of gene sets with q-val < 0.05 | GSEA Percentage of gene sets with q-val < 0.05 | SEA Percentage of gene sets with q-val < 0.25 | GSEA Percentage of gene sets with q-val < 0.25 |
|---|---|---|---|---|---|
| 1 | Pancreas | 20.1 | 0.0248 | 46 | 1.49 |
| 2 | P53 | 11.5 | 0 | 31.6 | 0.688 |
| 3 | Gender | 12.1 | 0 | 29.9 | 0.888 |
| 4 | Downs Syndrome | 13.3 | 0 | 34.3 | 0.488 |
| 5 | Cyclin D1 | 5.68 | 0 | 17.5 | 0.0904 |
| 6 | Rhabdoid | 19.8 | 0.005 | 46.1 | 1.01 |
| 7 | Revlimid Response | 17.3 | 0 | 42.4 | 0.501 |
| 8 | Metastasis | 18.8 | 0.0295 | 43.5 | 1.93 |
| 9 | Medulloblastoma monosomy 6 | 18.3 | 0.00641 | 43.6 | 2.28 |
| 10 | Lung Cancer KRAS | 11.4 | 0 | 32 | 0.39 |
| 11 | GSE6962 | 1.79 | 0 | 6.89 | 0.207 |
| 12 | GSE5462 | 26.5 | 0.0596 | 54.4 | 3.09 |
| 13 | Yeoh et al dataset | 2.85 | 0 | 9.42 | 0.0259 |
| 14 | GSE14615 | 22.4 | 0 | 49.1 | 1.41 |
| 15 | GSE7670 | 36.5 | 0.0132 | 66.4 | 3.25 |
| 16 | GSE14642 | 12.2 | 0 | 29.9 | 0.887 |
| 17 | GSE10961 | 27.4 | 0.0273 | 54.1 | 1.97 |
| 18 | GSE10281 | 21 | 0.0159 | 45.4 | 2.2 |
| 19 | Lung cancer survival | 16.6 | 0 | 40.1 | 0.738 |
| 20 | GSE10174 | 19.4 | 0.014 | 43.7 | 1.53 |
| 21 | GSE4475 | 27.8 | 0.00337 | 56.6 | 1.49 |
| 22 | GDS2668 | 9.98 | 0 | 24.3 | 0.0466 |
| 23 | Ovarian | 23.7 | 0.0184 | 49.6 | 2.02 |
| 24 | GDS495 | 6.42 | 0 | 17.4 | 0.0494 |
| 25 | GSE7562 | 4.69 | 0 | 15.1 | 1.11 |
| 26 | Melanoma | 13.9 | 0.00528 | 34.7 | 1.01 |
| 27 | GDS1306 | 10.6 | 0.0108 | 29.9 | 0.736 |
| 28 | GSE1413 | 4.27 | 0 | 15.8 | 0.168 |
| 29 | GSE9566 | 19.3 | 0.148 | 44 | 5.2 |
| 30 | Diabetes | 3.65 | 0 | 13.9 | 0.476 |
| 31 | Mouse lung cancer | 5.37 | 0 | 18.5 | 0.0852 |
| 32 | Trifluoperazine | 11.4 | 0 | 29.3 | 0.974 |
| 33 | Pioglitazone | 6.47 | 0.0142 | 18.9 | 0.944 |
| 34 | Vorinostat | 11 | 0 | 29.2 | 1.17 |
| 35 | Rosiglitazone | 10.2 | 0 | 26.1 | 1.31 |
| 36 | Troglitazone | 10.8 | 0 | 27.5 | 1.53 |
| 37 | Fluphenazine | 11.2 | 0 | 28.4 | 1.18 |
| 38 | Genistein | 10.7 | 0 | 26.9 | 1.28 |
| 39 | Alvespimycin | 9.65 | 0.00987 | 25.1 | 0.954 |
| 40 | Acetylsalicylic | 9.89 | 0.00578 | 25.9 | 0.938 |
| 41 | Alpha estradiol | 10.7 | 0 | 27.1 | 1.22 |
| 42 | Nordihydroguaiaretic | 10.1 | 0 | 26.4 | 0.885 |
| 43 | SHH | 4.89 | 0.00174 | 15.7 | 0.187 |
| 44 | GSE7403 | 12.4 | 0.00825 | 30.8 | 0.283 |
| 45 | GDS2025 | 18 | 0.0553 | 39.2 | 3.4 |
| 46 | SNF5 | 19.6 | 0 | 45.8 | 0.522 |
| 47 | HER2 | 12.1 | 0.00176 | 32.3 | 0.795 |
| 48 | Leukemia | 7.72 | 0 | 22 | 0.308 |
| 49 | Medulloblastoma histology | 2.23 | 0 | 9.41 | 0.106 |